\documentstyle[12pt]{article}
\begin{document}
\newcommand{\ket}[1] {\mbox{$ \vert #1 \rangle $}}
\newcommand{\bra}[1] {\mbox{$ \langle #1 \vert $}}
\newcommand{\bkn}[1] {\mbox{$ < #1 > $}}
\newcommand{\bk}[1] {\mbox{$ \langle #1 \rangle $}}
\newcommand{\scal}[2]{\mbox{$ \langle #1 \vert #2  \rangle $}}
\newcommand{\expect}[3] {\mbox{$ \bra{#1} #2 \ket{#3} $}}
\newcommand{\ki}{\mbox{$ \ket{\psi_i} $}}
\newcommand{\bi}{\mbox{$ \bra{\psi_i} $}}
\newcommand{\p} \prime
\newcommand{\e} \epsilon
\newcommand{\la} \lambda
\newcommand{\om} \omega   \newcommand{\Om} \Omega
\newcommand{\cc}{\mbox{$\cal C $}}
\newcommand{\w} {\hbox{ weak }}
\newcommand{\al} \alpha
\newcommand{\bt} \beta

\newcommand{\be} {\begin{equation}}
\newcommand{\ee} {\end{equation}}
\newcommand{\ba} {\begin{eqnarray}}
\newcommand{\ea} {\end{eqnarray}}

\def\lrD{\mathrel{{\cal D}\kern-1.em\raise1.75ex\hbox{$\leftrightarrow$}}}

\def\lr #1{\mathrel{#1\kern-1.25em\raise1.75ex\hbox{$\leftrightarrow$}}}
%i\! \lr{\partial_V}

\overfullrule=0pt \def\sqr#1#2{{\vcenter{\vbox{\hrule height.#2pt
          \hbox{\vrule width.#2pt height#1pt \kern#1pt
           \vrule width.#2pt}
           \hrule height.#2pt}}}}
\def\square{\mathchoice\sqr68\sqr68\sqr{4.2}6\sqr{3}6} 
\def\lrpartial{\mathrel
{\partial\kern-.75em\raise1.75ex\hbox{$\leftrightarrow$}}}

\begin{flushright}
%Tours\\
March 4th, 1997
\end{flushright}
\vskip 1. truecm
\vskip 1. truecm
\centerline{\LARGE\bf{The interpretation of the solutions}}
\vskip 2 truemm
\centerline{\LARGE\bf{of the Wheeler De Witt equation}}
%\centerline{\LARGE\bf{in Quantum Cosmology}}
\vskip 1. truecm
\vskip 1. truecm

\centerline{{\bf R. Parentani}}
\vskip 5 truemm

%\footnote{present address: 
\centerline{
Laboratoire de Math\'ematique et Physique Th\'eorique,
Facult\' e des Sciences}
%${ \quad\quad\quad\quad\quad\quad
%\quad\quad\quad } $  
\centerline{Universit\'e de Tours, 37200 Tours, France}

%\vskip 5 truemm
%\centerline{Laboratoire de Physique Th\'eorique de l'\' Ecole
%Normale Sup\'erieure\footnote{Unit\'e propre de recherche du C.N.R.S.
%associ\'ee \`a l'ENS et \` a l'Universit\'e de Paris Sud.}}
%%Ecole
%%Normale Sup\'erieure
%\centerline{24 rue Lhomond,
%75.231 Paris CEDEX 05, France.}
\centerline{
e-mail: parenta@celfi.phys.univ-tours.fr}
\vskip 5 truemm
\vskip 1.5 truecm
\vskip 1.5 truecm
\vskip 1.5 truecm
\vskip 1.5 truecm
%\vskip 1.5 truecm
{\bf Abstract }\\
%By treating matter interactions perturbatively, w
We 
%first 
extract transition
amplitudes among matter constituents of the universe
from the solutions of the Wheeler De Witt equation.  
%We then use 
%this treatment to determine how to
%interpret these solutions.
%the properties of these amplitudes to reach 
%perturbation theory constructed from
%these solutions 
%%designate 
The physical interpretation of these solutions
%wave function of the universe. 
is then reached by an analysis
of the properties of the transition amplitudes.
The interpretation so obtained is 
based on the current carried by these solutions 
and confirms ideas put forward
 by Vilenkin.

\vfill \newpage

\section{Introduction}

In the hamiltonian formulation of General Relativity
coupled to matter fields, the Euler-Lagrange equations
split into 6 dynamical equations plus the equations for
the matter fields and 4 constraints which express the
invariance of the theory under local changes of coordinates.
%reparametrizations
Therefore, when quantizing this system in a gauge 
invariant manner, i.e. \`a la Dirac,  the ``wave function''
must be annihilated by these four constraints that are
now operator valued.
This clear structure dictated by the reparametrization invariance
turns out to lead to very complicated problems concerning the
interpretation of this ``wave function''.
% governing matter and
%gravity configurations\cite{isham}\cite{hartle}\cite{Hal}\cite{kiefer2}\cite{vil}.
By interpretation, one designates the problem of extracting 
%from this wave 
(probabilistic) predictions concerning the evolution
of matter and gravity configurations.
% of matter and gravity.
The origin of the difficulties can be blamed on the absence
of time and the accompanying unitary evolution on which
quantum mechanics is based\cite{isham}\cite{hartle}\cite{Hal}\cite{kiefer2}\cite{vil}.

In order to describe more precisely some of these difficulties,
from now on we shall pursue the discussion in mini-superspace.
This drastic restriction to homogeneous and isotropic
three-geometries offers the double advantage of removing the
U.V. problem that plagues the local theory while keeping
the problem concerning the interpretation of the wave function.
In this restricted configuration space, 
%the sole gravitational degree
%of freedom 
gravity is described by the scale factor $a$, matter by homogeneous
fields that we shall denoted collectively by $\phi$ and the wave
$\Psi(a, \phi)$ is constrained to satisfy a single global
Wheeler De Witt (WDW) equation.
The questions 
%concerning  $\Psi(a, \phi)$ 
are the following:
How to proceed to read from $\Psi(a, \phi)$ predictions concerning 
quantum events such as,
for instance, transitions rates among the $\phi$
fields ?
%This question leads immediately to the next one:
and: Are those transitions described by a unitary evolution ?

Both questions have received attention and
many different schemes 
%on how to deal with $ \Psi(a, \phi)$
have been proposed\cite{isham}. 
%We refer to \cite{isham} for
%a list of these propositions. 
Let us mention here only the scheme based on the
hypothesis that $\vert \Psi(a, \phi) \vert^2$ does posses
a probabilistic interpretation\cite{hartle} (at least in terms of conditional
probabilities\cite{isham}) 
and the scheme based on the conserved current\cite{vil}
$\Psi^*i\!\!\lr{\partial_{a}}\!\! \Psi$. 
The answers to the second question are
partially related with this choice 
%and  illustrating the absence of 
%consensus, 
%are the answers to the second question. 
and range from, ``Yes, the evolution is unitary\cite{vent}\cite{kim}'',
to ``no, unitarity is violated\cite{kiefer2}'', 
and includes the middle attitude: ``unitarity is only approximatively conserved
\cite{vil}''.
The peculiar aspect of these widely distributed answers is that they
arise from the same starting point: 
%aris an agreement on the nature of 
the WDW equation and its solutions.
The disagreements build up with the choice of the
treatment 
%and the tools
 required to extract information from $\Psi(a, \phi)$. 

In the present article, we clarify the mathematical 
aspects of these treatments by analysing matter
interactions from the solutions of the WDW equation.
More precisely, by studying transitions, we identify the coefficient
${\cal{C}}_n(a)$ that weights the n-th state at $a$ 
and that replaces the amplitude $c_n(t)$ to be in the n-th 
matter state at time $t$
in conventional quantum mechanics. 
%and extract from and study the transitions they induce. 
The unambiguous identification is based on two criteria:
1) In the absence of transitions, ${\cal{C}}_n(a)$ must be constant.
 2) When one simplifies the equation governing their dependence in $a$
by treating gravity in the background field
approximation (BFA), the resulting equation must be the
 Schroedinger equation.
% evaluated in the gravitational background. 
This mathematical procedure of extracting 
%mathematically 
the ${\cal{C}}_n(a)$ 
is based on \cite{wdwgf}\cite{wdwpt}\cite{time}
% We stress
%that it 
and does not require an {\it a priori} interpretation of $\Psi ( a, \phi )$.
On the contrary, our program is first to determine 
the properties of the coefficients ${\cal{C}}_n(a)$
%when one abandons the BFA for describing gravity 
and only then to examine the question of its interpretation 
%of the coefficients ${\cal{C}}_n(a)$ 
in the light of these
%ir mathematical 
properties.
% of the coefficients ${\cal{C}}_n(a)$.
%Therefore, we shall first describe these properties with particular
%emphasis of the nature of the
%modifications induced by the abandonment of the
%background field approximation for describing the propagation of $a$.

They
%se properties 
reveal the existence of three regimes
which are delineated by the values of the
parameters describing matter transitions in quantum cosmology.
%a universe treated quantum mechanically.
In the case of weak interactions occurring close to
equilibrium in a macroscopic\cite{vil}\cite{BV}, 
the coefficients ${\cal{C}}_n(a)$ are equal to the
amplitudes $c_n(t)$, solutions of the corresponding 
Schroedinger equation evaluated in the geometry described by
$a(t)$.
In the second regime, the 
departure from equilibrium and/or the 
importance of the
interactions lead to ${\cal{C}}_n(a)$ that no longer coincides
with $c_n(t)$. However when gravity is still correctly described by 
WKB waves, the  ${\cal{C}}_n(a)$ still satisfy 
the ``unitary'' equation $\sum_n \vert {\cal{C}}_n(a) \vert^2 = 1$,
up to negligeable corrections.
%) even though their evolution can be 
%different that the one delivered by the Schroedinger equation.
In the third regime, the interactions are so violent that
the propagation of $a$ is affected by the quantum transition acts.
In that case, $\sum_n \vert {\cal{C}}_n(a) \vert^2 \neq 1$.
This ``violation'' is a direct manifestation 
of the modification of the propagation of $a$ by the
transitions themselves. It is
% corrections.
%We shall prove that
%the non-conservation of the ${\cal{C}}_n(a)$
%they are 
also kinematically related to the conservation
of the current carried by  $\Psi ( a, \phi )$.
%\Psi^*( a, \phi)\ i\!\!\lr{\partial_{a}} \Psi ( a, \phi )$
Under these extreme conditions, 
there is no possibility of defining a background.
Neither, therefore, should there be any possibility
of interpreting $\Psi (a, \phi )$ using the conventional rules
of quantum mechanics. This does not mean that no predictions
can be made, it simply means that the conventional
analysis cannot be performed {\it{in situ.}}
 By 
%``waiting'' enough that the universe leaves
propagating $\Psi (a, \phi )$
outside this regime, one can then perfectly determine its 
physical outcome.
%since they are determined by the properties of the solution there.

%We recall that the mathematical structure of perturbation theory
%constructed from the solution of the WDW equation shows
%that it is necessary to know the propagation of $a$ in order to 
%define ``to be in the n-th state at $a$'' since a reduction
%formula should be implemented, see \cite{wdwpt}. 
%Therefore there is no possibility to define
%a background
%%, this procedure becomes ambiguous. 
%Accordingly 
%%the definition of 
%the probability ``to be in the n-th state at $a$''
%also becomes ambiguous.
%We believe that any attempt to further analyse this
%problem, e.g. in terms of outcomes of a measurement described
%a la von Neeman, will confirm the physical emptiness to 
%define unambiguously this probability.

In conclusion, in this article, we shall show
%these properties lead to the following facts concerning 
%the interpretation of $\Psi (a, \phi )$ in the light of these properties.
%In particular, we emphasize\\ 
% that emerges from these properties is the following.
%Our analysis is incompatible with the interpretation based on
%conditional probabilities defined from the norm of $\Psi (a, \phi )$.
% On the contrary, it naturally fits into the scheme based  
%on its current. Thus, i
that when one requires that the conventional description 
of matter transitions is recovered from
quantum cosmology,
the probabilistic interpretation of  $\Psi (a, \phi )$ must be based 
on its current and not on its norm.
We shall also show that the interpretation of ${\cal{C}}_n(a)$ as the amplitude
of probability to find the n-th matter state at $a$
is valid as long as the propagation of gravity
is not significantly affected by the interactions.
Both aspects have been put forward by Vilenkin in \cite{vil}.
However, to the knowledge of the author, they have never been
made as explicit as in the present paper. Furthermore,
in contradistinction to \cite{vil},
our small parameter is the coupling constant among the quantum systems
and not their energy.
This allows to reach more general conclusions.

\section{The identification, the evolution and the meaning 
of the coefficients ${\cal{C}}_n(a)$}

As said in the Introduction, we shall use perturbation theory applied 
to matter interactions as a guide to identify the coefficients
${\cal{C}}_n(a)$. Before accomplishing this program, 
%one should first
we briefly present the kinematical properties at work in Quantum Cosmology,
in the absence of these interactions, see \cite{wdwgf} for more details.

\vskip.5 truecm
{\bf{The coefficients in absence of interactions}}
\vskip.2 truecm
\noindent
%In what follows, we briefly review the kinematical aspects
%of matter and gravity propagation in Quantum Cosmology
%that we have described at length in
%\cite{wdwgf}.
%here,  
%with them might consult 
%that reference wherein the same notations have been used.
For simplicity, the matter system is chosen in such a 
way that the free hamiltonian $H^{0}_m(a)$ does not
lead to transitions. Thus,  the general solution of the Schroedinger equation 
is 
\be
\chi(t, \phi) = \sum_n c_n e^{-i\!\int^t dt' E_n(t')}  \scal{\phi}{{n}}
\label{onee}
\ee
where the eigenstates of the  free hamiltonian $H^{0}_m(a)$ 
satisfy
\be
H^{0}_m(a(t)) \ket{n} = E_n(a(t)) \ket{n} 
\label{one}
\ee
Their time dependence arises only through the 
equation governing the background 
propagation $a = a(t)$.

The coefficients $c_n$ are interpreted as the amplitudes
of probability to find the matter system in the n-th state.
%since $\chi(t, \phi)$ gives the amplitude of probability to
%find the configuration $\phi$ at $t$.
Therefore, by convention, they are
normalized so that $\sum_n 
\vert c_n \vert^2 =1$. Furthermore, in the present case, they are constant. 
In this respect, notice that one must consider either interactions
with the external world, or self interactions, or non adiabatic 
transitions\cite{wdwpc} in order to 
give physical substance to 
the probabilistic interpretation of
$c_n$.
% itself (rather than its norm). 
Indeed, in the absence of interactions,
no interference among the $c_n$ will show up.

We shall now determine to what
extend these 
%well established 
properties are
%fully 
recovered in quantum cosmology from the solutions of the
Wheeler De Witt equation.

In minisuperspace, when matter is characterized by an
energy $E_n(a)$, the gravitational part of the action 
satisfies the Hamilton-Jacobi constraint
\ba
H_G(a) + E_n(a)=
% &=&0 \nonumber\\
{ -G^2  ( \partial_a S_G (a) )^2
% ; |A_M |^2,  E_\gamma ) 
 + \kappa a^2 + \Lambda a^4 \over 2 Ga } +
 E_n(a) =0
\label{M12}
\ea
where $G$ is Newton's constant, $\kappa$
is equal to $\pm 1$ or $0$ for respectively open,
closed and flat three surfaces and $\Lambda $ is the
cosmological constant. The
solution of this equation
%he gravitational action 
is simply $S_n (a)
= \int^a da' p_n(a')$ where the momentum of $a$
driven by the $E_n(a)$
%n-th matter energy is
\be
p_n(a) = \mp G^{-1} \sqrt{ 
\kappa a^2 + \Lambda a^4 + 2 Ga E_n(a) }
\label{M8}
\ee
where $\mp $ correspond respectively to expanding and
contracting universes.

Upon quantizing gravity, the Hamilton-Jacobi constraint
becomes the operator valued equation (the WDW equation)
\be
\left[
H_G(a) + H^0_m(a , \phi) \right] \Psi ( a, \phi ) =0
\label{wdw}
\ee
When the quantum matter states are characterized 
by constants of motion,
%, as in the present case, 
$\Psi ( a, \phi )$ can be decomposed as
\be
\Psi ( a, \phi ) = \sum_{n} {\cal{C}}_{n}
\Psi ( a; n ) \scal{\phi}{{n}}
\label{M13}
\ee
where the n-th gravitational wave, 
entangled\footnote{
This is the main reason for which we have chosen to work with 
matter states characterized 
by constants of motion. It allows an unambiguous 
decomposition of the total wave in terms of products. This is 
to be compared with the difficulties 
to perform this decomposition in the general case wherein
no clear principle seems to exist,
see e.g. \cite{kiefer2} after eq. (2.35) ``We {\it{choose}} $D$ in such a way that
the equations become simple''.
Very important also is the fact that our decomposition
 keeps the {\it linearity} of the WDW equation when used
 in a perturbative treatment.} to the n-th matter state, is a solution of 
\be
\left[
G^2 
%\hbar^2 
\partial^2 _a  
+ \kappa a^2 + \Lambda a^4 + 2 Ga E_n (a) \right]
 \Psi ( a; n ) =0
\label{M14}
\ee
This equation is second order in $\partial _a $ and has
therefore two independent solutions. 
This has to be the case since classically we can work
either with expanding or contracting universes.
%This choice is clearly exhibited 
Indeed when using the WKB approximation 
%wherein $\Psi ( a, n ) $ reads
\be
 \Psi ( a; n ) = { e^{ i\!\int^{a}
p_n(a')  da' } \over \sqrt{2 \vert p_n(a) \vert }}
\label{M144}
\ee
 one verifies that solutions with positive (negative)
wronskian
\be
 W_n = \Psi^*( a; n)\ 
i\!\!\lr{\partial_{a}} \Psi ( a;n ) 
\label{M1444}
\ee
correspond to expanding (contracting)
universes.
% in that semiclassical limit.

At this stage, several remarks should be made.
 Firstly, the decomposition of the general
solution of the WDW equation is performed by using the
same set of quantum numbers $n$ that the one
used in eq. (\ref{one}) wherein gravity was treated in the background
field approximation (BFA). The enlargement of the dynamics
to $a$ is compensated by the WDW constraint.
Secondly, it is now through the sign of the
Wronskian rather than at the classical level 
%in eq. (\ref{one}) 
that one now chooses expanding or contracting universes.
% is of course
%external to the Schroedinger equation.
Thirdly, with our  definition of 
$\Psi ( a;n )$, the coefficients ${\cal{C}}_{n}$ are
% by construction 
constant.

Having recall these kinematical properties, we can
now formulate precisely our mathematical claim\cite{wdwgf}: 
in order for ${\cal{C}}_{n} $
to satisfy the 
%usual time dependent 
Schroedinger equation
%in the limit in which 
when gravity is treated in the BFA, 
all Wronskians  $W_n$
must be equal to the same constant.
This can be already guessed by considering the 
{conserved} current\cite{vil} carried by $\Psi( a, \phi)$
\ba
 J = \int d\phi \ \Psi^*( a, \phi)
i\!\!\lr{\partial_{a}}  \Psi( a, \phi)
%\nonumber \\
= \sum_{n} \vert {\cal{C}}_{n} \vert^2 \ W_n
\label{curr}
\ea
To work with both $J=1$ and unit Wronskians suggests
an identification
of ${\cal{C}}_{n}$ with $c_n$.
However, at this moment, no physically relevant
conclusion\footnote{
It is nevertheless interesting to compare the present treatment based
on the current $J$ to the one in which it is the norm
%$\Psi^*( a, \phi)\Psi( a, \phi)$ 
that determines probabilities.
In that latter case, by definition, the probability to be in the 
n-th state is given by $\vert \scal{n}{\Psi( a, \phi)}
\vert^2 /  \scal{\Psi}{\Psi} = \vert {\cal{C}}_{n} \vert^2 p^{-1}_n(a)
/ \sum_{m} \vert {\cal{C}}_{m} \vert^2
/p^{-1}_m(a)$, where we have used the WKB form for the
waves $\Psi( a ;n )$.  This ``probability'' depends on $a$ 
through the $a$-dependent norm of each $\Psi( a; n )$,
%spread in $n$ and p_n(a)$, 
a feature that I find unattractive. Notice
however, that in the doubled limit of {\it well grouped} states
living in a  {\it macroscopic} universe, this dependence in $a$ vanishes, 
see latter in the text for
more precision concerning these limits.} can be
made concerning a probabilistic interpretation of the ${\cal{C}}_{n}$.
Indeed it is mandatory
to consider self interactions leading to transitions
%amongst the matter states 
%in order to determine how ${\cal{C}}_{n}$ evolve
%because, in the absence of interactions, the various 
in order for the ${\cal{C}}_{n}$ 
%components labeled by $n$ 
to vary 
and interfere, see the remark made after eq. (\ref{one}).

\vskip.5truecm
{\bf{The $a$-dependence of the ${\cal{C}}_{n}$}}
\vskip.2truecm
\noindent
We return for a moment to quantum mechanics
and consider the time
dependent perturbation theory
for allowing
a comparison with the corresponding equations
derived in quantum cosmology.
Upon introducing an interacting 
%(hermitian)
hamiltonian $H_{int}$ possessing non vanishing matrix
elements $\expect{n}{H_{int}}{m}$, the time dependence of the
coefficients
$c_n$ is given by 
\ba
i \partial_t c_n(t) = \sum_m
\expect{n}{H_{int}}{m}\ c_m(t) \ 
e^{-i \int^t dt' [E_m(t') - E_n(t')]}
\label{schr}
\ea
Since this equation is linear in $c_n$ and first order in $i \partial_t$,
when the hamiltonian $H_{int}$ is hermitian, one obtains
%immediately that \ba
$\Sigma_n \vert c_{n}(t) \vert^2 = 1$,
%\label{onene}
%\ea
 a necessary condition to keep the probabilistic interpretation of 
 $\vert c_n(t)\vert^2$.

In quantum cosmology, the fact of taking into account 
 the  
%(hermitian)
hamiltonian $H_{int}$ 
is quite different from what we just did in usual
quantum mechanics wherein one has an external
time parameter at our disposal.
 Indeed, 
%in quantum cosmology, the  interacting 
%(hermitian)
%hamiltonian 
$H_{int}$ modifies the propagation of
both gravity and matter through the modified WDW equation 
given by
\be
\left[
H_G(a) + H^0_m(a , \phi)   + H_{int}(a , \phi)  
 \right] \Xi ( a, \phi ) =0
\label{wdw'}
\ee
By decomposing the interacting wave $\Xi ( a, \phi ) $
in terms of the free components $\Psi ( a; n)$ 
\be
\Xi ( a, \phi ) = \sum_{m} {\cal{C}}_{m}(a) \
\Psi ( a; m ) \scal{\phi}{{m}}
\label{M13'}
\ee
and by projecting eq. (\ref{wdw'}) into the bra $\bra{n}$,
we obtain\footnote{Notice that this development does not coincide 
with the Born-Oppenheimer treatment. What would be
closer to that treatment, would consist in working with
 states which diagonalise the total hamiltonian
$H^0_m(a , \phi)   + H_{int}(a , \phi)$ at fixed $a$. 
Together with S. Massar, we shall present this adiabatic 
treatment applied to quantum cosmology in \cite{wdwpc}.}
%the equation determining the $a$-dependence of the ${\cal{C}}_{n}(a)$:
\be
 {\partial_a \Psi ( a; n ) \over \Psi ( a ; n )}
\partial_a {\cal{C}}_{n}(a) + {1 \over 2} \partial^2_a {\cal{C}}_{n}(a) 
+ {a \over G} \sum_{m} {\cal{C}}_{m}(a) \ \expect{n}{H_{int}}{m}
{ \Psi ( a ; m )\over \Psi ( a ; n )} = 0
\label{wdwg}
\ee
%This equation is linear in ${\cal{C}}_{n}$ and second order in to eq. (\ref{}), this 
%As eq. (\ref{}), 
Since this equation is 
second order in $\partial_a$, some analysis is required 
in order to reveal the properties of the evolution it encodes.
%(which is {\it{linear}} in ${\cal{C}}_{n}(a)$), 
To this end,
we shall first simplify it by making use of three approximations.
In the next subsection, we shall evaluate the errors they induce
on the basis of the analysis of \cite{wdwgf}\cite{wdwpt}.
 
The first approximation consists in using 
the WKB approximation for $\Psi ( a; n)$. 
Its validity requires $\partial_a p_n(a) / p^2_n(a) <\!\!< 1$, 
an inequality which
%a condition always 
is satisfied when the second condition is met.
This second condition requires that the universe be 
macroscopic \cite{vil}\cite{BV},
% character of the universe,
i.e. that the matter sources driving gravity must be macroscopic.
%the macroscopic character of the matter sources driving gravity.
By denoting $M$ the rest mass of the atoms and 
$\Delta m$ the energy change induced by the transitions
%states = E_{n'} - E_n$
%the energy scale characteristizing the microscopic transitions 
engendered by $H_{int}$, the macroscopic limit 
%character of the universe
guarantees that $\Delta m /E_n \simeq \Delta m/ N_M M <\!\!< 1$
since $N_M$, the total number of atoms, satisfies $N_M >\!\!>1$.
Thirdly, we 
%use the weak interaction limit necessary to
%validate the Golden Rule formula. By introducing the
require that the dimensionless constant $g^2$ that characterizes
the transition rates be smaller or comparable
to unity ($g$ is related to $H_{int}$ 
 by $\expect{n}{H_{int}}{m} \simeq g \Delta m$).
% This  in order for the transition rate 
%governing the transitions
%leads to a rate
%scales like $g^2 \Delta m$ and
%to satisfy $ g^2 \Delta m \ssimeq  \Delta m$.

By applying these three approximations to eq. (\ref{wdwg}),
one can drop the second term and write the two
other terms as
%and by keeping only the dominant terms, we obtain
\be
%i a G \sqrt{p_n(a) p_m(a)} \ \partial_a {\cal{C}}_{n}(a) =  
%\sum_{m}  \expect{n}{H_{int}}{m}\
i \partial_a {\cal{C}}_{n}(a) = \sum_{m} 
{a \over G}{\expect{n}{H_{int}}{m} \over \sqrt{p_n(a) p_m(a)}}\
{\cal{C}}_{m}(a) \ e^{i\!\int^a da'[p_m(a')- p_n(a') ] }\
{\sqrt{W_m  \over W_n}}
\label{niceeq}
\ee
This equation
% is the central equation of the paper.
%It should be compared with eq. (\ref{schr}) and
deserves a few comments.

Firstly, as eqs. (\ref{wdw'}, \ref{wdwg}), 
it is linear in ${\cal{C}}_{n}(a)$. 
We point out this fact since
%there exist in the literature 
many approximation schemes\cite{kiefer2}\cite{vent}\cite{kim}\cite{BV},
 as the semi-classical treatment,
%treatments of the solutions of the WDW equation which 
destroy the linearity of the WDW equation and therefore
the superposition principle. The loss
of linearity in these treatments results
from a quantum averaging performed at an earlier stage. 
 %in particular the semi-classical treatment does it, 
%see \cite{kiefer2}\cite{kim}\cite{vent}.

Secondly, in order for eq. (\ref{niceeq}) to coincide with 
%Schroedinger equation 
eq. (\ref{schr}) in the background
field approximation, the Wronskians must satisfy ${W_m / W_n} =1$
for all $m, n$. The validity of the further simplification which consists in treating 
eq. (\ref{niceeq}) in the BFA, requires that the ${\cal{C}}_{n}(a) \neq 0$
be grouped together such that their spread around the mean energy
$E_{\bar n}$ satisfies $(E_n - E_{\bar n})/E_{\bar n} <\!\!< 1$\cite{vil}\cite{BV}.
Only then can one correctly describe the evolution in terms of a 
single
%(proper\cite{wdwgf}) 
time parameter\cite{banks} defined, from the propagation of
 $a$, by
\be
 t_{\bar n}(a) = \int^a_{a_0} da' {a' \over G p_{\bar n}(a')}
\label{ta}
\ee
and develop the $a$-dependent phase of eq. (\ref{niceeq})
to first order in $E_m - E_n$ around $E_{\bar n}$, see \cite{wdwpt}. 
%In that case, u
Using
eq. (\ref{M8}) and $t_{\bar n}(a) $,
% given in eq. (\ref{ta}), 
one finds, for an expanding universe, 
\be
-\int^a_{a_0} da'(p_m(a')- p_n(a')) = 
%\int^a da' {a'
%\over G P(a'; \bar n) }
%(E_m(a')- E_n(a') )
 \int_0^{t_{\bar n}(a)} dt' (E_m(t')- E_n(t') ) + O((E_m - E_n)/ E_{\bar n})
\label{ta2}
\ee
%by definition of $t_{\bar n}(a) $ given in eq. (\ref{ta}).

Thirdly, when ${W_m / W_n} =1$, eq. (\ref{niceeq}) leads
to ``unitary'' evolution in the sense that $\Sigma_n \vert
{\cal{C}}_{n}(a) \vert^2 = 1$. We emphasize that this does not imply
that, starting with $ {\cal{C}}_{n}$ that coincide with $c_n$
at $a_0$ ($t=0),  
{\cal{C}}_{n}(a)$ will evolve like $c_n(t(a))$. Indeed, as
shown in eq. (\ref{ta2}), the 
%$a$-dependent 
phases of eq. (\ref{niceeq}) equal the corresponding phases 
of eq. (\ref{schr}) only when developed to first order in $\Delta E $ 
and evaluated for the mean energy $E_{\bar n}$. 
Therefore, 
%after a certain lapse of $t(a)$, 
the non-linear terms will induce increasing additional phase
shifts.
% that eventually will reach $\pi$. 
(This is not particular to quantum cosmology.
Indeed, whenever the quantum dynamics is enlarged to a variable formerly
treated classically, non-linear phase shifts appear,
%ence of these non-linear phases is generic,
see \cite{rec}.)
Then, after a certain lapse of $t(a)$, the interferences amongst
the ${\cal{C}}_{n}(a)$ will posses no relation to those amongst the $c_n(t(a))$.
%Moreover, in general, these additional phases lead to decoherence. 
Moreover, remote configurations evolve with their own time\cite{vil}, 
see eq. (\ref{ta}) for the dependence on $E_n$ in $t_n$.

Fourthly, eq. (\ref{niceeq}) might have been obtained by
``first quantizing'' the wave $\Xi(a, \phi)$. In that framework,
one
postulates that the fundamental equation 
%that determines this wave 
is
\be
-i \partial_a \Xi(a, \phi) = \sqrt{ H_0 (a, \phi) + H_{int}(a, \phi)} \ \Xi(a, \phi) 
\label{firstq}
\ee
instead of eq. (\ref{wdw'}), see \cite{isham}. 
Then, by developing the square root to first order in $ H_{int}$
one obtains eq. (\ref{niceeq}) exactly like eq. (\ref{schr}) is obtained\cite{wdwpt}.
%i.e. by first defining free states.
The main weakness of this {\it ad hoc} approach is that there is no a priori
justification to eliminate half of the solutions of the Hamilton Jacobi
equation before quantization. Furthermore, whether or not it offers a good 
approximation, the importance of the neglected terms cannot be estimated
without considering the solutions to eq. (\ref{wdw'}). 

%The important point is that there is a regime in Quantum Cosmology,
%in which the evolution does not coincide with the corresponding one
%evaluated at the BFA but which still leads to $\Sigma_n \vert
%{\cal{C}}_{n}(a) \vert^2 = 1$.

Finally, the interpretation
of the wave $\Xi ( a, \phi )$ as the conditional amplitude
of probability to find $\phi$ at $a$ is rejected by the present 
analysis 
%based
% the interpretation
%of the wave $\Xi ( a, \phi )$ as the conditional amplitude
%of probability to find $\phi$ at $a$ 
of the 
solutions of  eq. (\ref{wdw'}). 
%(Notice that this is not the case if one
%works with the solutions of  eq. (\ref{firstq}).)
%
Indeed, by adopting this interpretation,
one would obtain a non-linear equation
%for the dynamical equation 
for the conditional amplitudes 
since these are non linearly related to ${\cal{C}}_{n}(a)$ 
--recall the presence
of the normalisation factor $p_n^{-1/2}(a)$ 
stemming from current conservation, see the second footnote--. 
%From a linear dynamical equation, the WDW equation,
%It is unnatural to propose an interpretation of the 
%solutions of a linear equation in terms
%of amplitudes which evolve non linearly. 
Notice that this would not have been the case if one
would have worked with the solutions of eq. (\ref{firstq}).
Notice also that, in contrast to what is presented in 
\cite{Hal}, our derivation of eqs. (\ref{wdwg}, \ref{niceeq})
which encode the correlations between matter and gravity 
in no way requires that $\Xi(a, \phi)$ be peaked around certain
configurations. But it does require that gravity be modified
by the transitions, otherwise $a$ could not be used to parametrize
their amplitudes.  
%In this we differ from \cite{hartle}\cite{Hal} since, 
%in these works, the physical predictions are based on the peaks
%of  $\Xi(a, \phi)$.

We now address the problem of the corrections to eq. (\ref{niceeq}).
%and to their implications to $\Sigma_n \vert
%{\cal{C}}_{n}(a) \vert^2$.

\vskip .5 truecm
{\bf The  $a$-dependence of $\Sigma_n \vert
{\cal{C}}_{n}(a) \vert^2$}
\vskip .2 truecm
\noindent
Our aim is to determine how the terms neglected in passing from
eq. (\ref{wdw'}) to eq. (\ref{niceeq}) affect $\Sigma_n \vert
{\cal{C}}_{n}(a) \vert^2$.
%To this end, it is sufficient to determine the dominant departure from
%``unitarity''. 
This is most easily achieved by using the {\it exact} conservation
law of the current carried by $\Xi ( a, \phi )$.
%since it is then sufficient to 
%use only eq. (\ref{niceeq}).
Indeed the current carried by $\Xi ( a, \phi )$ is 
\ba
J &=& \int d\phi \left( \Xi^*( a, \phi)
i\!\!\lr{\partial_{a}}  \Xi( a, \phi) \right)
\nonumber \\
&=& \sum_{n} \vert {\cal{C}}_{n} \vert^2  + \sum_{n} \left( {\cal{C}}_{n}^* (a)
i\!\!\lr{\partial_{a}} {\cal{C}}_{n} (a)\right)
\vert \Psi (a; n) \vert^2 = 1
\label{curr'}
\ea
We first notice that there is no
systematic departure from unity. Indeed, when 
the interactions cease to act, due for instance to some 
cooling in an expanding universe, the final value of $\Sigma_n \vert
{\cal{C}}_{n}(a) \vert^2 $ coincides with the initial one. 

Before discussing the reasons that lead to the 
local departure from unity,
%$\Sigma_n \vert {\cal{C}}_{n}(a) \vert^2 \neq 1$, 
we estimate the importance of this departure. To this end,
it is sufficient to use eq. (\ref{niceeq}) and the
WKB expression for the free waves $\Psi(a, \phi)$. One should also 
 further specify the characteristics of the solution under
examination. As an example, in the case of a matter dominated
universe  composed of $N_M$ ``heavy'' atoms of mass $M$ 
having transitions between inner states characterized by 
energy gaps $\Delta m$,
this correction is
%obtains
\ba
\sum_{n} \left( {\cal{C}}_{n}^* 
i\!\!\lr{\partial_{a}} {\cal{C}}_{n} \right)
\vert \Psi (a; n) \vert^2 \simeq
2 \sum_{n} \sum_{m} {\cal{C}}_{m } {\cal{C}}_{n}^* {a \expect{n}{H_{int}}{m}
\over G p_n(a) p_m(a)}
\simeq g {a N_M^{1/2 }\Delta m \over G p^2_n(a)}
 \simeq g { \Delta m \over M } {1 \over N_M^{1/2 } }
\label{depart}
\ea
far from a turning point\cite{km} and at equilibrium. 
(The incoherence of the transitions
close or at  equilibrium brings the factor 
$N_M^{1/2 }$ in the second approximation.)
The departure from unity is therefore small for three independent reasons: the
weakness of the transitions rates controlled by $g$, the
smallness of the $\Delta m / M$ which allows to neglect the recoil
of the atoms induced by the transitions and the macroscopic character
of the universe $ N_M >\!\!> 1$. 

The mechanism that leads to the 
 departure from unity is clear: when the ${\cal{C}}_{n}$
depend on $a$, they carry some current and therefore their norm
should vary accordingly.
Indeed, when the ``potential term'' of
a second order differential equation varies, 
the induced modifications of the norm and the phase of the solutions 
are correlated since the Wronskian is conserved.
%In order to understand t
The physical origin of 
these correlated modifications
%, notice that
%The origin of $\Sigma_n \vert
%{\cal{C}}_{n}(a) \vert^2 \neq 1$ can  
comes from the fact that the WDW imposes 
that the kinetic energy of gravity be 
modified by the presence of $H_{int}$, see
eq. (\ref{wdw'}). 
This in turn modifies the norm of the solution.
Therefore, $\Sigma_n \vert {\cal{C}}_{n}(a) \vert^2 \neq 1$ 
is a manifestation of the modification of the 
propagation of gravity induced by the interactions
themselves.

To obtain explicitly the relation between the
changes in phase and in amplitude, it is instructive
to consider the simple case wherein the interactions
induce a diagonal level shift $E_n(a) \to E_n(a) + \delta_n(a)$.
Using eq. (\ref{M8}) in eq. (\ref{M144}), one immediately gets the desired
relation through the change of the $p_n(a)$. 
% between the changes in phase and amplitude.
Upon considering matter transitions, this modification of   $p_n(a)$
has the following physical meaning.
The transition probability is given by an integral over $a$ whose 
norm is $a da/p_n(a)=dt_n$, see \cite{wdwpt}. Therefore,
the modification of $p_n(a)$ is
{\it necessary} in that it guarantees 
%appear upon evaluating the derivative 
%of the transition amplitudes is 
 that the Golden Rule is obtained, i.e. that
the transition probability increases linearly with the proper time lapse,
eq. (\ref{ta}), 
evaluated in the universe wherein the transition occurs.
%of $p_n(a)$ when evaluated in terms of the proper time,
%see \cite{wdwpt}. 

In view of this, it is inviting to work with eigenstates of 
the total matter hamiltonian $H_0 + H_{int}$ rather than to work
in perturbation theory with the free states. Indeed, when working with 
these eigenstates, 
%of the total hamiltonian, 
one automatically includes
the backreaction of the (adiabatic part of the)
interaction hamiltonian in the definition of 
the WKB momentum $p_n(a)$.
% on the same footing that we did
%it with the free hamiltonian. 
Since the induced modification of $p_n(a)$
 will no longer be found, its contribution
to $\Sigma_n \vert {\cal{C}}_{n}(a) \vert^2 \neq 1$ will be also
absent. 
This strongly suggests not to interpret
$\Sigma_n \vert {\cal{C}}_{n}(a) \vert^2 \neq 1$ 
as a violation of unitarity since the magnitude of the
departure from unity depends
on the perturbative scheme adopted. 
Together with S. Massar, we shall return to these aspects
in \cite{wdwpc}.

\vskip .4 truecm
{\bf Conclusions and additional remarks}
\vskip .1 truecm
\noindent
In resume, our analysis of ${\cal{C}}_{n}(a)$ shows that there are three
regimes delineated by the values of the coupling constant $g$, the relative
transition energy $\Delta m/M$ and the macroscopic character of the universe,
here controlled by $N_M$.

In order to be in the first regime, the initial values ${\cal{C}}_{n}(a_0) \neq 0$
must be grouped together so that the ``mean'' time parameter, eq. (\ref{ta}),
 correctly parametrizes the 
evolution of the whole system.
One must also requires that the universe be macroscopic
and that the transitions be not too violent. 
In that regime, ${\cal{C}}_{n}(a) = c_n(t_{\bar n}
(a))$. Therefore, ${\cal{C}}_{n}(a)$ {\it is} the amplitude of probability to 
find the matter system in the n-th state at $a$.

In the second regime, the spread of ${\cal{C}}_{n}(a_0) \neq 0$
and/or the violence of the interactions and/or the ``smallness'' of the universe
leads to an evolution that cannot be obtained from a Schroedinger equation
based on a single time parameter.  Nevertheless, when the 
WKB approximation for describing the free 
evolution of gravity is still valid, one still obtains a linear equation
for the propagation of the ${\cal{C}}_{n}(a)$
that guarantees that $\Sigma_n \vert {\cal{C}}_{n}(a) \vert^2 = 1$.
Moreover, upon considering a set of neighbouring ${\cal{C}}_{n}(a)$ for
a sufficiently small amount of time, the evolution of this restricted set
still obeys a Schroedinger equation. 
Finally, as usually in quantum mechanics, remote configurations do not interfere.
Therefore, it is still perfectly correct to interpret
${\cal{C}}_{n}(a)$ as the amplitude of probability to 
find the n-th state at $a$.
Notice however that the decoherence of the background time, that is the
fact that remote configurations evolve with their own time, would leave no
physical meaning to mean values of matter operators in which
non-interfering configurations would contribute. Only summations 
over restricted sets of interfering configurations make physical sense.
In this we differ from \cite{hartle}\cite{Hal} since, 
in these works, physical predictions are based on the peaks
of  $\Xi(a, \phi)$ which include summation over all states.

In the third regime, the importance of the interactions leads to
backreaction effects on the propagation of gravity such that 
$\Sigma_n \vert {\cal{C}}_{n}(a) \vert^2$ depends on $a$.
A part of this dependence comes from the (adiabatic\cite{wdwpc}) dressing energy 
brought in by the interactions. It reflects the fact that
the relationship between proper time lapse and the momentum $p(a)$
has been modified by this dressing energy. Therefore the deviation from unity
it engenders should certainly not be interpreted as a violation of unitarity. 
This deviation can be also understood from the necessity of applying a ``reduction 
formula'' to the universe's wave function in order to obtain 
transition amplitudes amongst its constituents, see \cite{wdwpt}.

Far more difficult to interpret are the consequences of the
corrections to the WKB approximation. 
Indeed upon dealing with exact solutions for 
the propagation of $a$,  initially expanding (forward)
solutions contain, at other radii, backward waves
corresponding to contracting universes.
Moreover, the conservation of the Wronskian tells
us that the current of the forward part has increased.
The only way to confront this ``Klein paradox'' seems
to proceed to the so called third quantization\cite{isham}.

However, we want to emphasize that by adopting 
this framework, one has not solve the question of extracting transition amplitudes
occurring {\it within} expanding universes.
% since this framework does
%not designate a single procedure for doing so.
%extracting transition amplitudes
%occurring within a specific universe. 
Indeed, additional choices
must be specified in order to know how to extract from  superpositions
of expanding and contracting universes predictions concerning 
transitions in the expanding sector. 
%We believe that these additional choices will be rather disconnected to the 
Without having further specified these choices and 
having considered combinations of forward and backward solutions,
 it is overhasty to conclude that unitarity will (not) be violated in Quantum 
Cosmology.

\vskip 1. truecm
\vskip .5 truecm

{\bf Acknowlegdments}

\noindent
I am indebted to R. Brout for the clarifying discussions 
that we had upon writing the review article\cite{time}
wherein similar problems are addressed.
%the aspects developed in this paper, as well as for useful
%comments concerning a first draft of this paper.
I am also indebted to S. Massar for helpful remarks
provided during the last stage of the writing.
Finally, I wish to thank Y. Aharonov, Cl. Bouchiat, A. Casher, Fr. Englert, 
J. Iliopoulos and T. Jacobson
for useful discussions.

\end{document}